\documentclass[12pt]{iopart}
\usepackage{braket} 
\begin{document}

\title[Time-extended measurement of the position]{Time-extended measurement of the position of a driven harmonic oscillator}

\author{Francesc S Roig}

\address{Physics Department, University of California, Santa Barbara, CA 93106, USA}
\ead{roig@physics.ucsb.edu}
\begin{abstract}
The von Neumann interaction between a particle and an apparatus, both of arbitrary mass, has been considered in the measurement of the position of a simple harmonic oscillator acted on by an external force. When the measurement has finite duration, both the motion of the pointer and the oscillator influence the result of the measurement. Provided that the oscillator is in an eigenstate of its position at the start of the measurement, the pointer will indicate the arithmetic average between the initial and final position of the particle with an added term which depends on the duration of the measurement and the frequency of the oscillator. This additional term is determined by the external force which also causes the appearance of a phase factor in the wave function at the end of the measurement. This phase factor depends on the average of the initial and final positions of the particle. Furthermore, the probability that the pointer indicator variable will correlate with a given average value is equal to the transition probability for the undisturbed free oscillator to experience the change in position. If the initial state of the pointer is a narrow wavepacket, then for any initial state of the oscillator, the measurement yields, approximately, the undisturbed probability distribution for the position of the free oscillator at the end of the measurement. The transition probability for the pointer to experience a change in its position has also been evaluated.

\end{abstract}

\maketitle

\section{Introduction}
The current work is an extension and generalization of the article in~\cite{FR} by the present author.  That work provided a detailed analysis of  the finite time  measurement of the position of a free particle when the measuring apparatus and the particle have finite masses.  The case of a simple harmonic oscillator interacting with an apparatus of infinite mass was also considered in detail. 

We consider a harmonic oscillator acted on by an external force interacting with an apparatus of finite mass. The Hamiltonian for this system is
\begin{equation} \label{H}
H=\frac{p^2}{2m}+V(x,t)+\frac{P^2}{2M}+H_{\rmi}.
\end{equation}
The mass of the oscillator is~$m$ and the mass of the apparatus or pointer is~$M$. The total potential acting on the oscillator of frequency~$\omega$ is
\begin{equation} \label{V(x,t)}
V(x,t)=V(x)-f_{\rm D}(t)x,
\end{equation}
 where~$V(x)=\frac{1}{2}m\omega^2x^2$~and~$f_{\rm D}(t)$ is the external or driving force acting on the oscillator. 

The interaction between the particle and the apparatus is described by the von Neumann Hamiltonian~\cite{vonN} for the measurement of the position of a particle
\begin{equation} \label{Hi}
H_{\rmi} = \frac{1}{T} f(t) xP,
\end{equation}
where~$P$ is the momentum of the pointer,~$T$ the duration of this interaction, and $f(t)$ is a dimensionless coupling function of time with compact support [0,T].  Just before the measurement, the system is described by the pure state $\psi_0(x, X) = \varphi_0(x) \Phi_0(X)$, where $\varphi_0(x)$ is the state of the particle and~$\Phi_0(X)$ can be a narrow wavepacket describing the apparatus, with~$X$ the pointer position, or indicator variable.  This wavepacket can be  centered at~$X=0$. If we assume that $T$ is very short, compared to any dynamical time scale for particle and pointer, then the interaction~(\ref{Hi}) is very large compared to all the other terms in (\ref{H}) and by itself makes the system evolve to the entangled state at the end of the measurement
\begin{equation} \label{mEnd}
\psi(x, X) = \varphi_0(x) \Phi_0(X - g x),
\end{equation}
where
\begin{equation} \label{g}
g = \frac{1}{T} \int_0^T f(t)\rmd t
\end{equation}
is a dimensionless coupling constant. The final state~(\ref{mEnd}) of the system is the product of the state of the particle, unaffected by the measurement, and the state of the pointer, which correlates with the position of the particle. For a measurement that is very fast the wave packet for the pointer has not spread, and its center has been displaced by an amount given by the shift function $s(x) = g x$. Furthermore, the probability distribution that in the final state of the system the indicator variable correlates with a value~$x$ of the position of the particle is 
\begin{equation}\ \label{P(x)}
P(x) = \int_{-\infty}^\infty \left|\varphi_0(x) \Phi_0(X - g x)\right|^2\rmd X.
\end{equation}
For a normalized pointer state $\Phi_0(X)$, this is
\begin{equation} \label{AbsP}
P(x) = \left|\varphi_0(x)\right|^2,
\end{equation}
 the same as the probability distribution for the position in the initial state of the particle.
This is in agreement with the general characterization of measurement theory in~\cite{LEBall}: The probability distribution, that in the final state of the system the apparatus indicates the value~$x$ is~$\left|\varphi_0(x)\right|^2$, just the same as the probability distribution for the position of the particle in the state before the measurement. It is in this sense that this is a perfect measurement.  In a measurement in finite time the kinetic energy of the particle cannot be neglected and the position of the oscillator is no longer a constant of the motion, unlike the case with the interaction~(\ref{Hi}) alone; thus the measurement will be imperfect~\cite{Araki-Yanase,Ghirardi-et-al}.  That is, the distribution of the position of the particle in the state just before the measurement cannot be obtained from the state of the system when the measurement has been completed.

In this paper, we will consider the case of any coupling function~$f(t)$ in~(\ref{Hi}) and a driving force~$f_{\rm D}(t)$, both symmetric about the midpoint of the interval~$[0,T]$.  We will find that for a driven oscillator in an eigenstate of its position~$x$ at~$t=0$, at the end of the measurement at time~$T$ the shift function for the pointer is given by an expression of the form
\begin{equation} \label{sh}
s(x,x')=g(\omega,T)\left(\frac{x+x'}{2}\right) + d(\omega,T),
\end{equation}
where~$\omega$ is the frequency of the oscillator,~$x$ and~$x'$ are the initial and the final position of the oscillator,~$g(\omega,T)$ is a dimensionless  function of the frequency and the  time~$T$, and the displacement term~$d(\omega,T)$ appears as a result of the driving force.   That is, the pointer correlates with the arithmetic average of the initial and final position of the oscillator, and an additional constant term is added, when compared with the result obtained in~\cite{FR}, to yield the position of the pointer at the end of the measurement.  If either the coupling function or the driving force, or both, are not symmetrical about the midpoint of the time duration of the measurement, then the shift function consists of a linear combination of the initial and final positions of the oscillator plus a constant term which again is due to the presence of the external force. Finite time von Neumann measurements were considered first in~\cite{Peres/Wooters} for the spin of a particle, and more recently in~\cite{FR} for the case of the position of a particle. The present work describes a single measurement that takes a finite amount of time. This is  to be distinguished from the continuous monitoring of the position of a particle as in~\cite{MenskyI,AV,Caves,Caves/Milburn}.  The continuous observation of the position of an oscillator acted on by an external force was considered in~\cite{Mensky}.  A formalism for measuring the time average of any dynamical quantity on individual Feynman histories is developed in~\cite{Sokolovski,Liu/Sokolovski} . Also see~\cite{Sokolovski/Mayato} for the general relation between restricted paths sums and von-Neumann-like quantum measurements.  

A final note regarding the meaning of a quantum {\it{measurement}}: In this article we adopt the view from~\cite{Peres/Wooters} regarding the distinction between a quantum {\it{ measurement}} and an {\it{observation}}. We characterize a quantum measurement as a correlation between the value of the position variable of the particle with the states of the apparatus after the measurement, whereas an observation is the process of selection of a particular value of, in this case, the position of the pointer and the position of the particle, which happens through the collapse of the wavefunction.  For measurements of finite duration the correlation between particle and pointer takes place through the arithmetic value of the initial and final position of the particle, and this generates the entanglement between the two. In the current work we treat the particle and the apparatus as quantum subsystems that interact with each other, and their evolution is entirely governed by the Schr\"odinger equation.  This is to be distinguished from hybrid models of measurement~\cite{Peres-Terno,Chua,Elze}, where some degrees of freedom are treated quantum mechanically and others are treated classically. That is, there is a {\it{cut}}~\cite{Hay-Peres} between the quantum and classical worlds.  

Furthermore, the system under consideration in our work is exactly soluble.  The solution to the finite time quantum measurement problem for the driven oscillator contains, as particular cases, the exact solution for a free particle, a particle acted on by a a time-dependent force, and a free harmonic oscillator.  In addition, the pointer can have finite or infinite mass.  In some cases in the literature~\cite{Caves,Caves/Milburn,Sokolovski,Liu/Sokolovski,Sokolovski/Mayato} explicitly or implicitly it is assumed that the apparatus has infinite mass. Thus the influence of the motion of the pointer during the measurement is ignored.

Apart from the intrinsic interest in quantum measurement theory when a particle is moving in a quadratic potential, the main motivation for the current paper is its relevance in the generalized quantum mechanics of closed systems~\cite{HartleQC,Hartle}, particularly in the study of decoherence issues when the closed system consists of a driven oscillator and an apparatus that performs a finite time quantum measurement of the position of the oscillator.  That is, a measurement situation~\cite{Gell-Mann/Hartle,HartleQC} ensues. The formalism of quantum mechanics of closed systems requires the system to be treated quantum mechanically, and in this approach there is no cut between the quantum world and the classical world.  These issues will be treated in a forthcoming work by this author where the decoherence of coarse-grained histories in spacetime is explored.

The model described by~(\ref{H}-\ref{Hi}) is developed in Sec. II of this paper, and its behavior for a pointer with finite mass is found in Sec. III. In Sec. IV we study the probability distribution for the position of the particle when the measurement is completed as well as its relation to the different transition probabilities for the oscillator and for the pointer.


\section{Formalism. The propagator for the system oscillator-apparatus}
In this section we will develop the basic formalism for the evaluation of the propagator for the Hamiltonian~(\ref{H}).  In~\cite{FR} it was shown that the propagator for a particle-pointer system, with~$\hbar=1$,  is written as a sum over all paths between~$0$ and~$T$ as
\begin{equation} \label{pathint}
\left\langle {x,X} \right|\rme^ { - \rmi HT} \left| {x',X'} \right\rangle  = \!\!\int \!\!\!\int \delta x(t)\delta X(t)\exp\left\{\rmi S[x(t),X(t)] \right\},
\end{equation}
where the action is
\begin{equation}\label{S}
S\left[ {x(t),X(t)} \right] = \int_0^T\left[ \frac{m}{2}\dot x^2  - V(x,t)
+\frac{M}{2}\left( {\dot X - \frac{f(t)}{T}x} \right)^2\right] \rmd t.
\end{equation}

Inserting two complete sets of eigenstates of the momentum~$P$ of the pointer, normalized according to~$
\left\langle P \right|\left. {P'} \right\rangle  = \delta (P - P')$,
 the propagator can be rewritten
\begin{eqnarray} \label{propH_P}
\left\langle {x,X} \right|\rme^{ - \rmi HT} \left| {x',X'} \right\rangle  =& \int_{ - \infty }^\infty  {\frac{\rmd P}{2\pi }} \exp{\left\{\rmi \left[ P(X - X') - \frac{P^2}{2M}T\right]\right\}}\nonumber\\
 &\times\left\langle x \right|\exp{( - \rmi H_{\rm P} T)} \left| {x'} \right\rangle,
\end{eqnarray}
where
\begin{equation} \label{H_P}
H_{\rm P}=\frac{p^2}{2m}+V(x,t)+\frac{f(t)}{T}xP.
\end{equation}
The reduced propagator~$\left\langle x \right|\exp {( - \rmi H_{\rm P} T)} \left| {x'} \right\rangle$ for the particle can be expressed as a sum over all particle paths and the propagator in~(\ref{propH_P}) becomes
\begin{eqnarray} \label{reduced_path_int}
\left\langle {x,X} \right|\rme^ { - \rmi HT} \left| {x',X'} \right\rangle  =& \int_{ - \infty }^\infty  \frac{\rmd P}{2\pi } \exp{\left\{ \rmi\left[ P(X - X') -\frac{P^2}{2M}T\right]\right\}}\nonumber\\
&\times \int \delta x(t) \exp\left\{\rmi S_{\rm P}[x(t)]\right\},
\end{eqnarray}
where the reduced action is
 \begin{equation} \label{S_P}
S_{\rm P}[x(t)]=\int_0^T{L_{\rm P}}\rmd t
\end{equation}
and~$L_P$ is the reduced Lagrangian
\begin{equation} \label{L_P}
L_{\rm P}=\frac{m}{2}\dot x^2-V(x,t)-\frac{f(t)}{T}xP.
\end{equation}
For a driven oscillator the reduced Lagrangian is
\begin{equation} \label{L_PF}
L_{\rm P}=\frac{m}{2}\dot x^2-\frac{m}{2}\omega^2 x^2+F(t,P)x
\end{equation}
with
\begin{equation} \label{F(t,P)}
F(t,P)=f_{\rm D}(t)-\frac{f(t)}{T}P.
\end{equation}

The Lagrangian~(\ref{L_PF}) is quadratic and therefore the path integral in~(\ref{reduced_path_int}) is determined by the classical action for the reduced Lagrangian
\begin{equation} \label{path_int=S_cl}
\int {\delta x(t)\exp\left\{\rmi S_{\rm P} [x(t)]\right\} }  = A_{\rm{mp}}\exp{\left[\rmi S_{\rm{cl}}(P)\right]}, 
\end{equation}
where~$A_{\rm{mp}}$ is an amplitude factor and~$S_{\rm{cl}}(P)$ is the classical action.
If both~$f(t)$ and~$f_{\rm D}(t)$ are symmetric about the midpoint of the measurement interval~$[0,T]$, then using a familiar result from~\cite{Feynman} the classical action for the reduced Lagrangian~(\ref{L_PF}) is
\begin{eqnarray} \label{Sclasical}
S_{\rm{cl}}(P)  =&\frac{m\omega}{2\sin\omega T}[(x^2\!+\!x'^2)\cos\omega T\!-\!2xx']
\!+ \!\frac{(x + x')}{\sin \omega T}\int_0^T \!\!{F(t,P)\sin\omega t~\rmd t}\nonumber\\ 
&- \frac{1}{m \omega\sin \omega t}\int_0^T\! \!{\int_0^t \!\!{F(t,P)F(s,P)
\sin \omega (T - t)\sin \omega s}}~\rmd t\rmd s.
\end{eqnarray}
If either~$f(t)$ and~$f_{\rm D}(t)$, or both, are not symmetric about the midpoint of the measurement interval~$[0,T]$, then we would obtain that the coefficients of~$x$ and~$x'$ in the expression above are not the same. In this case the pointer will not indicate the arithmetic average of the initial and final positions of the particle, but instead it will correlate with a linear combination of the initial and final positions.\\
The amplitude factor is given by
\begin{equation} \label{Amp}
A_{\rm {mp}}=\left( \frac{m\omega}{2\pi \rmi \sin\omega T} \right)^{1/2}.
\end{equation}
Inserting~(\ref{F(t,P)}) in~(\ref{Sclasical}) the following expression for the classical action is obtained:
\begin{eqnarray} \label{SclP}
S_{\rm{cl}}(P)=&\frac{m\omega}{2\sin\omega T}[(x^2+x'^2)\cos\omega T-2xx']-\frac{A_{\rm D}(\omega,T)}{m\omega \sin\omega T}\nonumber\\
&+\frac{B_{\rm D}(\omega,T)}{\sin\omega T}\left(\frac{x+x'}{2}\right)
+\frac{PC_{\rm D}(\omega,T)}{ª\omega \sin\omega T}\nonumber\\
&-\frac{PB(\omega,T)}{T\sin\omega T}\left(\frac{x+x'}{2}\right)-\frac{P^2A(\omega,T)}{T^2m\omega\sin\omega T}
\end{eqnarray}
where
\begin{eqnarray} \label{A_D}
A_{\rm D}(\omega,T)=\int_0^T\int_0^t&\rmd t\rmd sf_{\rm D}(t)f_{\rm D}(s)\nonumber\\
&\times\sin\omega (T-t)\sin\omega s
\end{eqnarray}
\begin{equation} \label{B_D}
B_{\rm D}(\omega,T)=2\int_0^T\rmd t{f_{\rm D}(t)\sin\omega t}
\end{equation}
\begin{eqnarray} \label{C_D}
C_{\rm D}(\omega,T)=\int_0^T\int_0^t&\rmd t\rmd s[f_{\rm D}(t)f(s)+f(t)f_{\rm D}(s)]\nonumber\\
&\times\sin\omega (T-t) \sin \omega s
\end{eqnarray}
\begin{equation} \label{A}
A(\omega,T)=\int_0^T\int_0^t\rmd t\rmd s{f(t)f(s)\sin \omega (T-t)\sin\omega s}
\end{equation}
\begin{equation} \label{B}
B(\omega,T)=2\int_0^T\rmd t{f(t)\sin\omega t}.
\end{equation}
The propagator for the system oscillator-pointer is then
\begin{eqnarray} \label{propagator}
\left\langle {x,X} \right|e^{ - \rmi HT} \left| {x',X'} \right\rangle  =& \int_{ - \infty }^\infty  {\frac{\rmd P}{2\pi }} \exp{\left\{ \rmi\left[ P(X - X') - \frac{P^2 }{2M}T\right]\right\}}\nonumber\\
&\times
A\exp{[\rmi S_{\rm{cl}}(P)]},
\end{eqnarray}
where the amplitude~$A_{\rm {mp}}$ is given by~(\ref{Amp}) and~$S_{\rm{cl}}(P)$ is given by~(\ref{SclP}). 

Next we insert the classical action~(\ref{SclP}) into~(\ref{propagator}). The integration over the momentum of the pointer is easily carried out to obtain
\begin{eqnarray} \label{PropTOT}
 \left\langle {x,X}\right|\rme^{- \rmi HT} \left| {x',X'} \right\rangle =&K_{0}(x,T;x',0)\exp\left[\rmi  \phi\left(\frac{x+x'}{2},\omega,T\right)\right]  \left(\frac{M_\mathit{\rm {eff}} }{2\pi \rmi T}\right)^{1/2}\nonumber\\
&\times
\exp \left\{ {\rmi \frac{M_\mathit{\rm {eff}}}{2T}\left [ {X - X' - s(x,x')} \right ]^2} \right\},
\end{eqnarray}
where
\begin{equation}\label{s(x,x')}
s(x,x')=g(\omega,T)\left(\frac{x+x'}{2}\right)+d(\omega,T).
\end{equation}
The coupling constant multiplying the arithmetic average of the initial and final position in~(\ref{s(x,x')})  is given by
\begin{equation} \label{g_omegaT}
g(\omega,T)=\frac{B(\omega,T)}{T\sin\omega T},
\end{equation}
where~$B(\omega,T)$ is given by~(\ref{B}).\\
The effective mass is
\begin{equation} \label{Meff}
M_\mathit{\rm {eff}} = M\left[1 + \frac{2 A(\omega, T) M}{m \omega T^3 \sin(\omega T)}\right]^{-1},
\end{equation}
 and~$A(\omega,T)$ is given by~(\ref{A}).\\
The factor~$K_{0}$ is the propagator for the free harmonic oscillator:
\begin{eqnarray}\label{K_0}
K_0(x,T; x',0) =& \left( \frac{ m \omega }{2 \pi \rmi \sin \omega T} \right)^{ 1/2 }\!\!\! \exp \left\{\rmi\frac{  m\omega }{2 \sin \omega T }\right.\nonumber\\
& \times[ ( x^2 + x'^2 ) \cos \omega T - 2 x x']\left.\vphantom{\frac{m}{2}}\!\!\right\}.
\end{eqnarray}
The driving force introduces two terms into the propagator:  
a displacement term
\begin{equation} \label{d}
d(\omega,T)=-\frac{C_{\rm D}(\omega,T)}{m\omega T\sin\omega T}
\end{equation}
with~$C_{\rm D}$ determined by~(\ref{C_D}),
and a phase factor~$\exp(\rmi\phi)$ in~(\ref{PropTOT}), where the phase is given by 
\begin{equation} \label{Phi}
\phi =\frac{1}{\sin\omega T}\left[\left(\frac{x+x'}{2}\right)B_{\rm D}-\frac{A_{\rm D}}{m\omega}\right],
\end{equation}
with $A_{\rm D}$ and~$B_{\rm D}$ are determined by~(\ref{A_D}) and~(\ref{B_D}) respectively.
In the absence of the driving force the propagator for the system does not have  the displacement term~(\ref{d}) nor the factor with the phase~(\ref{Phi}).  The coupling constant~(\ref{g_omegaT}) and  the effective mass term~(\ref{Meff}) are the same as in the driven case.

In the limit of a very short measurement then from~(\ref{PropTOT}) the familiar  von Neumann result follows
\[
\Braket { x, X | \rme^{- \rmi  H T}| x', X'} \mathop  = \limits_{T \to 0}  \delta ( x - x' ) \delta \left( X - X' - f (0) x \right) ,\nonumber\\
\]
where~$f(0)$ is the dimensionless coupling function in~(\ref{Hi}) evaluated at~$t=0$.

Also, in the limit~$\omega\rightarrow0$, then~(\ref{PropTOT}) for the propagator of the system becomes the propagator for a particle acted on by a force~$f_{\rm D}(t)$ when both the force and the coupling function in~(\ref{Hi}) are symmetric about the midpoint of the measurement time interval.  The effective mass is now
\begin{equation} \label{Meff-omega=0}
M_\mathit{\rm {eff}}  = M\left( {1 + \frac{2M}{mT^4 }\!\int_0^T \!\!{\rmd t\int_0^t \!\!{\rmd s(T - t)sf(t)f(s)} } } \right)^{ - 1} 
\end{equation}
and the coupling constant~(\ref{g_omegaT}) becomes
\begin{equation}\label{g(T)}
g(T)=\frac{2}{T^2}\int_0^T{\!\!tf(t)}\rmd t.
\end{equation}\
The displacement~(\ref{d}) is now
\begin{equation} \label{d(T,0)}
d(T)=-\frac{1}{mT^2}\int_0^T\int_0^t[f_{\rm D}(t)f(s)+f(t)f_{\rm D}(s)]
(T-t)s\rmd t\rmd s.
\end{equation}
The phase~(\ref{Phi}) is written
\begin{equation} \label{Phi(T)}
\phi=(x+x')\int_0^T{f_{\rm D}(t)t}\rmd t 
- \frac{1}{mT}\int_0^T\int_0^t{f_{\rm D}(t)f_{\rm D}(s)(T-t)s}\rmd t\rmd s.
\end{equation}

In~\cite{FR}, based on the structure of (13) in that work, the factorization property of the propagator for quadratic potentials was conjectured: The propagator for the system is written as a product of two factors. The first factor is the propagator for the particle when the interaction with the apparatus is turned off.  The second factor is the propagator for a free particle of mass~$M_{\mathit{\rm{eff}}}$ with the pointer coordinate~$X$, where the initial position was shifted by an amount proportional to~$\bar x = \frac{(x+x')}{2}$.  The present work shows that the conjecture holds for free particles and free oscillators. For a driven oscillator the propagator of the system contains three factors:  namely, the propagator for the free oscillator; the propagator for a free particle of mass~(\ref{Meff}) with position coordinate~$X$, whose initial position contains a total shift consisting of a term proportional to~$\bar x$ plus a term given by~(\ref{d}); and an overall phase factor~$e^{\rmi\phi}$, with~$\phi$ given by (\ref{Phi}).  The presence of a linear factor in the Hamiltonian of the particle leads to the appearance of the phase factor and the additional coordinate independent shift.


\section{Oscillator-apparatus state after the measurement}
Unitary evolution will determine the state of the system oscillator-apparatus at the end of the measurement at time~$T$. That is,
\begin{equation} \label{Psi}
\psi (x,X,T) =\int_ {- \infty} ^\infty  {\int_{ - \infty }^\infty { \left\langle {x,X} \right|e^{ - \rmi HT} \left| {x',X'} \right\rangle } }
\psi _0 (x',X')\rmd x'\rmd X',
\end{equation}
where the initial state of the system is~$\psi_0(x,X)=\varphi_0(x)\Phi_0(X)$~ and~$H$ is the Hamiltonian~(\ref{H}), with the potential
\begin{equation}\label{V(x,t)SHO}
V(x,t)=\frac{m}{2}\omega^2x^2-f_{\rm D}(t)x.
\end{equation}
After inserting~(\ref{PropTOT}) into~(\ref{Psi}), we obtain
\begin{equation} \label{Psi(x,X,T)}
\psi(x,X,T)=\int_{-\infty}^{\infty}\rmd x'{\varphi_0(x')\psi_{\rm D}(x,x',X,T)}
\end{equation}
where the driving force~$f_{\rm D}(t)$ is contained in the term
\begin{eqnarray}\label{Psi_D}
\psi_{\rm D}(x,x',X,T)=&\exp\left [\rmi \phi\left(\frac{x+x'}{2},  \omega,T\right)\right ] K_0(x,T;x',0)\nonumber\\
&\times
\Phi_{\rm M_{\rm{eff}}}( X-s(x,x') ),
\end{eqnarray}
with the shift function 
\begin{equation}\label{eq:s(x,x')}
s(x,x')=g(\omega,T)\left(\frac{x+x'}{2}\right)+d(\omega,T),
\end{equation}
where~$g(\omega,T)$ and~$d(\omega,T)$ are given by~(\ref{g_omegaT}) and~(\ref{d}) respectively.\\
The factor~$K_0(x,T,x',0)$ in~(\ref{Psi_D}) is the propagator~(\ref{K_0}) for the free oscillator. The factor~$\Phi_{\mathit{M}_\mathit{eff}}$ is given by the expression
\begin{eqnarray}\label{PhiMeff}
\Phi_{ {\rm  M }_{\rm{eff} } } ( X -  s(  x,  x'  ) )= & \left ( \frac { { M }_\mathit { \rm{eff} } } {  2\pi \rmi T }\right )^{ 1/2 }\!\!\!\int_{- \infty }^{  \infty } \!\!\!\rmd X'  \Phi _0 ( X' ) \nonumber\\
&\times
\exp \left
\{\rmi \frac { M_{\rm{eff}} }{ 2 T } [X - X' - s( x,  x' )  ]^2 \right\},
\end{eqnarray}
which exhibits the entanglement between the oscillator and the pointer.
This expression can be rewritten in the form
\begin{eqnarray}
\Phi_{  \rm M _{\rm{eff} } } ( X -  s(  x,  x'  ) )= & \left ( \frac { { M }_\mathit { \rm{eff} } } {  2\pi \rmi T }\right )^{ 1/2 }\int_{- \infty }^{  \infty } \rmd X'  \Phi _0 ( X'-s(x,x') ) \nonumber\\
&\times
\exp \left
[ \rmi \frac { M_\mathit {\rm{eff}} }{ 2 T } (X - X' )^2 \right],
\end{eqnarray}
which describes a spreading wavepacket centered at~$X_0=s(x,x')$ at~$t=0$.  At time~$T$ this spread corresponds to the evolution of a free particle with mass~$M_{\mathit{\rm{eff}}}$. 

If the initial state of the oscillator is an eigenstate of the position with eigenvalue~$x_0$, then~$\varphi_0 (x) = \delta (x-x_0)$ and from~(\ref{Psi(x,X,T)}),~(\ref{Psi_D}) and~(\ref{PhiMeff})
 we obtain the wavefunction for the oscillator apparatus system at the end of the measurement
\begin{eqnarray}\label{Psi_Dx_0}
\psi (x , X, T) = &K_0 (x, T ; x_0, 0) \exp \left [ \rmi  \phi \left( \frac { x+x_0 }{ 2 }, \omega, T \right )\right ]\nonumber\\
&\times
\Phi_{\rm M_{\rm{eff}}}\left ( X-g(\omega,T)\left (\frac{x+x_0}{2}\right )-d(\omega,T)\right ).
\end{eqnarray}
This expression shows that after the measurement the pointer indicates the arithmetic average of the initial and final position of the oscillator. In addition a position-independent term~$d(\omega,T)$ has been added to the indication of the pointer at the end of the measurement. This added term exhibits the effect of the driving force on the measurement. The driving force also introduces a phase factor~$\rme^{\rmi \phi}$  in the final state wavefunction with the phase~$\phi$ given by~(\ref{Phi}) evaluated at the initial oscillator position~$x'=x_0$. 
When the initial state of the oscillator is not a sharp state of the position, then~(\ref{Psi(x,X,T)}) and~(\ref{Psi_D}) show that the final state is a superposition of states of the form~(\ref{Psi_Dx_0}).  If the range of the position of the oscillator in the initial state~$\varphi_0( x )$ is~$x_0-\Delta  < x < x_0+ \Delta$, then after the measurement is completed, for a given final position~$x$ of the oscillator the pointer has moved, and the possible values of the shift function~(\ref{s(x,x')}) spread continuously in the range~$(\bar x -  \Delta/2)g  <  s ( x, x' ) <  (\bar x + \Delta/2)g$, where~$g$ is given by~(\ref{g_omegaT}) and $\bar x$ is the arithmetic average of~$x_0$ and~$x$.


\section{The probability of the position of the oscillator in the state after the measurement}
 Next we consider the probability distribution of the position of the oscillator in the final state described by~(\ref{Psi(x,X,T)}) and~(\ref{Psi_D}).  
 We can rewrite~(\ref{PhiMeff})
\begin{equation}\label{PhiMeff2}
\Phi_{\rm M_{\rm{eff}}} \left ( X - s(x, x') \right ) = 
\int_{-\infty}^\infty \left\langle {X} \right|\hat U_{\rm{eff}}(T,0)
 \left|{X'+s(x,x')} \right\rangle
\left \langle {X} | \Phi_0 \right \rangle \rmd X',
\end{equation}
where $s(x,x')$ is the shift (\ref{s(x,x')}), and
\begin{equation}\label{Ueff}
\hat U_{\rm{eff}}(T,0) = \exp {\left(-\rmi \frac{\hat P^2 T}{2 M_{\rm {eff}}} \right)}
\end{equation}
is an effective time evolution operator for a free particle of mass~$M_\mathit{\rm{eff}}$,  with~$\hbar =1$ and~$\hat P$ the pointer momentum operator. The probability distribution for the position of the oscillator is obtained by integrating over the pointer coordinate 
\begin{eqnarray}\label{AbsPsi}
\int_{-\infty}^\infty \rmd X | \psi (x, X, T ) |^2 =& \int_{-\infty}^\infty \rmd X \bigg| \int_{-\infty}^\infty  \rmd x' 
\left\langle {x} \right|\hat U_0(T,0) \left| {x'} \right\rangle 
 \left \langle {x'}|\varphi_0 \right \rangle
\nonumber\\
&\times 
\int_{-\infty}^\infty  
\rmd X' \exp\left[ \rmi \phi \left(\frac{x+x'}{2}, \omega, T \right)\right]\nonumber\\
&\times
\left\langle {X} \right|\hat U_{\rm{eff}}(T,0)
\left| X'+s(x,x') \right\rangle\left \langle {X'}|\Phi_0\right \rangle \bigg|^2,
\end{eqnarray}
where~$\left\langle {x} \right|\hat U_0(T,0) \left| {x'} \right\rangle = K_0(x,T;x',0)$, the propagator~(\ref{K_0}) for the free oscillator, and~$s(x,x')$ is given by~(\ref{eq:s(x,x')}).\\
The integration over~$\rmd X$ in~(\ref{AbsPsi}) produces a~$\delta$-function kernel 
\begin{eqnarray}\label{delta_kernel}
&\int_{-\infty}^\infty \rmd X  \left\langle {X''+s(x,x'')} \right|\hat U_{\rm{eff}}^\dagger(T,0) 
\left| {X}\right\rangle\ 
\left\langle {X} \right|\hat U_{\rm{eff}}(T,0) 
\left| {X'+s(x,x')} \right\rangle\ \nonumber\\
&= \delta \left(X'-X''+g(\omega, T) \frac{x'-x"}{2}\right).
\end{eqnarray}
Next in~(\ref{AbsPsi})  the integrations over the pointer variables are collected, and after inserting~(\ref{delta_kernel}) we obtain
\begin{eqnarray}\label{X_integrations}
\int_{-\infty}^{\infty} & \rmd X
\int_{-\infty}^{\infty}\rmd X' \int_{-\infty}^{\infty}  
\rmd X''\left\langle {X''+s(x, x'')} \right|\hat U^\dagger_{\rm{eff}}(T,0) 
\left| {X} \right\rangle\nonumber\\
&\times
\left\langle {X} \right|\hat U_{\rm{eff}}(T,0) 
\left| {X'  +s(x,x')} \right\rangle
\left \langle {\Phi_0} |X'' \right \rangle 
\left \langle {X'} | \Phi_0 \right \rangle\nonumber\\
& =
\int_{-\infty}^{\infty}\rmd X'\Phi_0^\star\left(X' \!\! + g(\omega,T)\frac{x'-x''}{2}\right) \Phi_0(X'),
\end{eqnarray}.\\
Finally, the following expression is obtained for the probability distribution:
\begin{eqnarray}\label{AbsPsi_final}
\int_{-\infty}^\infty \rmd X | \psi (x, X, T ) |^2 =& \int_{-\infty}^\infty \rmd x''K_0^\star ( x, T; x'', 0 ) \varphi^\star_0 ( x'' )\nonumber\\\nonumber\\
& \times
\int_{-\infty}^\infty  \rmd x' K_0 (x, T; x' 0 ) \varphi_0 (x') \nonumber\\
& \times
\exp \left [ - \rmi  \phi \left( \frac{x+x''}{2}, \omega , T  \right) \right] \nonumber\\
&\times \exp \left [  \rmi  \phi \left( \frac{x+x'}{2}, \omega , T  \right) \right] 
\Delta_{\rm{overlap}},
\end{eqnarray}
with the pointer overlap factor
\begin{equation}\label{Delta_overlap}
\Delta_{\rm{overlap}} = \int_{-\infty}^\infty \Phi_0^\star \left ( X + g (\omega, T) \frac{x'-x''}{2} \right)
\Phi_0 (X) \rmd X.
\end{equation}
The entanglement between the oscillator and the apparatus takes place through this overlap integral.\\  
When~(\ref{AbsPsi_final}) is integrated  over the position of the oscillator, and the initial state is normalized, we obtain the result
\begin{equation}
\int_{-\infty}^\infty \int_{-\infty}^\infty \rmd x\rmd X\left|\psi(x,X,T)\right|^2 = 1,
\end{equation}
as expected by unitarity.\\
The total phase in the phase factors in~(\ref{AbsPsi_final}) can be worked out with the help of~(\ref{Phi}) 
\begin{equation}\label{phi-phi}
 \phi\left ( \frac { x+x' } { 2 },\omega, T \right ) - \phi \left ( \frac{x+x''}{2},\omega,T \right ) = \frac { B_{\rm D} }{ \sin \omega T } \left (\frac  {x'-x'' } { 2 } \right ),
\end{equation}
where~$B_{\rm D}$ is given by~(\ref{B_D}). The effect of the driving force only appears through this phase term.\\
The overlap integral in~(\ref{Delta_overlap}) can be expanded in powers of~$( x' - x'' )$ and written
\begin{eqnarray}\label{Phi_expand}
&\int_{-\infty}^\infty  \Phi_0^\star \left(  X + g ( \omega , T )\frac{ x' - x''}{2}\right) \Phi_0 ( X ) \rmd X  \nonumber\\
& \approx\int_{-\infty}^\infty \Phi_0^\star ( X ) \Phi_0 ( X ) \rmd X + O ( x' - x'' ).
\end{eqnarray}
If we assume that the initial state of the pointer is a very narrow normalized wavepacket, then we get a significant contribution to~(\ref{Delta_overlap}) only in the region~$x''\approx x'$, and we can use the approximation
 ~$\Delta_{\rm{overlap}} \approx 1$.  Likewise, the phase~(\ref{phi-phi}) vanishes in this approximation.\\
 Under these conditions the probability distribution for the position of the oscillator in the final state is approximately
\begin{equation}~\label{prob_at_T}
\int_{-\infty}^\infty \int_{-\infty}^\infty \rmd x \rmd X | \psi (x, X, T ) |^2 \approx |\varphi (x, T )|^2
\end{equation}
with
\begin{equation}\label{phi(x,T)}
\varphi (x , T ) = \int_{-\infty}^\infty \rmd x' K_0(x , T ; x' , 0 ) \varphi_0 (x') .
\end{equation}
Thus, if the initial state of the pointer is narrow, the entanglement between the oscillator and the pointer is negligible.
Furthermore, the probability distribution for the position of the oscillator in the state right after the measurement is approximately the same as the probability distribution in the state at time~$T$ corresponding to the oscillator evolving with no external force, and being uncoupled from the pointer. In short,  whether the oscillator is free or driven, the result in this case is the same with regard to the probability distribution after the measurement. The driving force only influences the displacement of the pointer at the end of the measurement. A similar result was also obtained in~\cite{FR} for the case of a free particle.  In this sense this is an almost perfect measurement of the position of the oscillator.   On the other hand, if the initial state of the pointer is not narrow, the entanglement particle-pointer is not negligible, and the influence of the phase difference~(\ref{phi-phi}) cannot be neglected either. The measurement is imperfect.  Next we will consider the two physical situations:  First, the initial state of the oscillator is an eigenstate of position and the initial state of the pointer is a normalized wavepacket.  Second the initial state of the pointer is an eigenstate of position and the initial state of the oscillator is normalizable.

 
\subsection{Sharp oscillator state at the start of the measurement}
The initial state of the system oscillator - pointer is 
\begin{equation}\label{sharp_oscillator_state}
\psi_0 (x , X , T ) = \delta (x - x_0 )\Phi_0(X).
\end{equation}
Then it follows from~(\ref{AbsPsi_final}) that the probability distribution is
\begin{equation}\label{prob-sharp-oscillator}
\int_{-\infty}^\infty \rmd X|\psi(x,X,T)|^2 = \left| K_0 (x , T; x_0 , 0 ) \right|^2
\end{equation}
or, substituting for the propagator of the free harmonic oscillator~(\ref{K_0}), we obtain the probability distribution
\begin{equation}\label{const_dist}
\int_{-\infty}^\infty \rmd X|\psi(x,X,T)|^2 = \frac{m\omega}{2\pi\sin\omega T}~.
\end{equation}
This is the same result  obtained in~\cite{FR} for a free harmonic oscillator and a pointer with infinite mass. 
The finite mass of the apparatus and the presence of a driving force do not influence this result.  The initial state of the system is not normalizable and therefore this result is consistent with the uniform relative probability distribution in~(\ref{const_dist}).

We can solve for the Heisenberg equations of motion for the driven oscillator:
\begin{eqnarray}\label{Heisenberg-x}
\hat x(t) =\hat x(0) \cos \omega t + \frac {\hat p(0)}{m \omega} \sin \omega t
 +
\int_0^t G( t, t' ) f_{\rm D}(t')\rmd t',
\end{eqnarray}
\begin{eqnarray}\label{Heisenberg-p}
\hat p(t) = && -m\omega \hat x(0) \sin \omega t + \hat p(0) \cos \omega t\nonumber\\
&&
-m\omega^2\int_0^t \rmd t'\int_0^{t'} \rmd t'' G(t' , t'') f_{\rm D}(t'') + \int_0^t \rmd t' f_{\rm D}(t'),\nonumber\\ 
\end{eqnarray}
and~$G(t,t')$ satisfies the equation
\begin{equation}\label{G}
\left (m \frac {\rmd ^2}{\rmd t^2} + m \omega^2 \right ) G( t, t' ) = \delta ( t - t' ),
\end{equation}
with~$G ( t, t' ) = 0$ for~$t < t'$.
In the position basis , and with~$ \hbar = 1$, the eigenstates of~$\hat x (T)$ are solutions of the equation
\begin{eqnarray}\label{eigeneqn}
 x(T) \Braket { x | x(T) }=& x\cos \omega T \Braket { x | x(T) } - \rmi \frac{\sin \omega T}{m \omega} \frac{\partial} {\partial x} \Braket { x | x(T) }\nonumber\\
&+ F_{\rm D}(T) \Braket { x | x(T) },
\end{eqnarray}
where
\begin{equation}\label{F(T)}
F_{\rm D}( T ) = \int_0^T G( T, t' ) f_{\rm D}(t')\rmd t'.
\end{equation}
The solution to (\ref{eigeneqn}) is
\begin{eqnarray}\label{soln-eigen}
\Braket { x | x(T) } = C\exp \bigg\{ -\rmi \frac { m \omega }{ \sin \omega T } \bigg[ \frac{ x^2 }{ 2 } \cos \omega T
+ x F_{\rm D}(T) - x(T) x  \bigg]  \bigg\},
 \end{eqnarray}
where~$F_{\rm D}(T)$ is given by~(\ref{F(T)}) and~$C$ is an arbitrary constant.\\
The~$\delta$ function normalization
\begin{equation}\label{delta}
\Braket { x'(T) | x,(T) } = \delta \left ( x'(T) - x(T)\right )
\end{equation}
yields the value
\begin{equation}\label{C}
C = \left [ \frac { m \omega }{ 2 \pi \sin \omega T } \right ]^{1/2}\!\!\!.
\end{equation}
The normalized state~(\ref{soln-eigen}) gives the transition probability for the oscillator to go from~$x(0)$ to~$x(T)$
\begin{equation}\label{trans-prob}
\left | \Braket { x(T) | x(0) } \right |^2 =  \frac { m \omega }{ 2 \pi \sin \omega T }. 
\end{equation}
This result is just~$\left |K_0(x,T;x_0,0)\right |^2$, the probability distribution for the oscillator at the end of the measurement.  That is,
\begin{equation}\label{tran-prob}
\left | \Braket { x(T) | x(0) } \right |^2 =\int_{-\infty}^\infty \rmd X |\psi (x , X, T )|^2.
\end{equation}
It easily follows from~(\ref{Heisenberg-x}) that the transition probability from~$x(0)$ to~$\bar x = [x(0) + x(T)]/2$ is given by
\begin{equation}\label{trans-prob2}
\left | \Braket { \bar x | x(0) } \right |^2 = 2 \int_{-\infty}^\infty \rmd X |\psi (x , X, T )|^2.
\end{equation}
Finally, if the average is taken over a Feynman path 
\begin{equation}\label{path-average}
\hat{\bar x} = \frac {1}{T} \int_0^T \hat x(t) \rmd t,
\end{equation}
then the following result is obtained for the transition probability
\begin{equation}\label{path-average2}
\left | \Braket { \bar x | x(0) } \right |^2 = \omega T\left[ \tan \left(\frac{\omega T}{2} \right)\right]^{-1} \int_{-\infty}^\infty \rmd X |\psi (x , X, T )|^2.
\end{equation}
Whether the oscillator is free or is acted on by an external force, results~(\ref{trans-prob}),~(\ref{trans-prob2}) and~(\ref{path-average2}) hold. The driving force does not influence these results.

 
\subsection{Sharp pointer state at the start of the measurement}
The initial state of the system oscillator - pointer is 
\begin{equation}\label{sharp-oscillator-state}
\psi_0 (x , X  ) =  \varphi_0(x) \delta(X).
\end{equation}
Inserting~$\Phi_0(X) = \delta (X)$ into~(\ref{Delta_overlap}) 
we obtain
\begin{equation}
\Delta_{\rm{overlap}}=\frac{2}{g(\omega,T)}\delta(x' - x''),
\end{equation}
and inserting this result into~(\ref{AbsPsi_final}) yields the probability distribution 
\begin{equation}\label{prob-final-result}
\int_{-\infty}^\infty \rmd X | \psi (x, X, T ) |^2 = \frac { m \omega }{ g( \omega, T ) \pi \sin \omega T},
\end{equation}
where~$g(\omega,T)$ is given by~(\ref{g_omegaT}).
This result is the same for all normalized initial states of the oscillator.  Furthermore, the relative probability distribution is uniform, which is consistent with a non-normalizable initial state. The driving force does not influence this result.

In the limit~$\omega\rightarrow 0$ we obtain the result valid both for a free particle and for a particle acted on by a force
\begin{equation}\label{omega=0}
\int_{-\infty}^\infty \rmd X | \psi (x, X, T ) |^2 = \frac{ m }{g( T)\pi T},
\end{equation}
with~$g(T)$ given by~(\ref{g(T)}).

We can solve for the Heisenberg equations of motion for the pointer to obtain
\begin{equation}\label{P(t)}
\hat P(t) = \hat P(0) = \hat P_0.
\end{equation}
The momentum of the pointer is a constant of the motion.
The position of the pointer evolves in time according to
\begin{equation}\label{X(t)}
\hat X(t) =  \hat X_0 + \frac{\hat P_0}{M} t + \frac{1}{T} \int_0^t f(t') \hat x(t') \rmd t'.
\end{equation}
In order make the evaluation of the transition probability for the pointer more tractable we let the coupling function in~(\ref{Hi}) be a dimensionless constant, that is~$f(t) = g$.  Next we insert~(\ref{Heisenberg-x}) into the previous expression for~$\hat X(t)$ and at time~$T$ we obtain
\begin{equation}\label{X(T)}
\hat X(T) = \hat X_0 + \frac{\hat P_0}{M} T +  \frac{g}{T}\left[\frac{2 \hat p_0}{m \omega ^2} \sin^2\frac{\omega T}{2} 
 + \frac{\hat x_0}{\omega} \sin \omega T + G_{\rm D}(T) \right],
 \end{equation}
 where
 \begin{equation}\label{G_D(T)}
G_{\rm D}(T) =  \int_0^T \rmd t' \int_0^{t'} \rmd t'' G( t' , t'' ) f_{\rm D}(t'').
\end{equation}
The equation satisfied by the eigenstates of the position of the pointer at time~$T$ with eigenvalue~ $X_{\rm T}$, 
\begin{equation}\label{eigenX}
\hat X(T)
\left| {X_{\rm T}} \right\rangle = X_{\rm T} 
\left| {X_{\rm T}} \right\rangle, 
\end{equation}
can be rewritten in terms of the position variables~$x$ and~$X$ of the oscillator and the pointer respectively at~$t = 0$, to obtain
\begin{eqnarray}\label{Psi(x,X)}
&\left[ X - X_{\rm T} + \frac{g x}{\omega T} \sin \omega T + \frac{g}{T} G_{\rm D}(T) \right]
\Psi _{\rm{X_T }}(x,X) = 
\nonumber\\
&\rmi \hbar\left[ \frac{2g}{T m \omega^2}\sin^2 \left(\frac{\omega T}{2}\right)\frac{\partial}{\partial x} + \frac{T}{M} \frac{\partial}{\partial X} \right ]\Psi_{\rm{X_T}}(x,X).
\end{eqnarray}
The solution to~(\ref{Psi(x,X)}) can be written as a product
\begin{equation}\label{phi-Phi}
\Psi_{\rm{X_T}}(x,X) = \varphi_{\rm T}(x) \Phi_{\rm{X_T}}(X),
\end{equation}
 and we readily obtain
\begin{equation}\label{phi_T(x)}
\varphi_T(x) = A_{\varphi}\exp\left\{-\rmi \frac{m\omega^2 T}{2 \hbar g}\left[ \sin\left(\frac{\omega T}{2}\right)\right]^{-2}
\left ( Cx + \frac{g x^2}{2 \omega T} \sin \omega T\right)\vphantom{\frac{\omega^2}{\sin^2\left(\frac{\omega T}{2}\right)}}\right\},
\end{equation}                                   
where~$A_{\varphi}$ and $C$ are constants.
This factor does not depend on the driving force, but only on quantities referring to the free oscillator.\\
The factor with the pointer variable is
\begin{equation}\label{Phi_X_T(X)}
\Phi_{\rm{X_T}}(X) = A_{\Phi} \exp \left \{\rmi \frac{ M}{\hbar T} \left [ X_T X - \frac{X^2}{2}
- \frac{g}{T} G_{\rm D}(T)X + C X \vphantom { \frac{ X^2 }{ 2 } } \right ] \vphantom{\frac{M}{T}}\!\right \}.
\end{equation}
This factor depends on the driving force through~$G_{\rm D}(T)$ defined in~(\ref{G_D(T)}) and exhibits the dependence on the position~$X_{\rm T}$ of the pointer at time~$T$.\\
Then it follows from~(\ref{phi-Phi}),~(\ref{phi_T(x)}) and~(\ref{Phi_X_T(X)})
\begin{equation}
\Braket { X'_{\rm T} |X_{\rm T} } =|A|^2\int_{-\infty}^{\infty}\rmd X\exp \left [\rmi \frac{M}{\hbar T}(X'_{\rm T} - X_{\rm T})X\right ],
\end{equation}
where~$A=A_{\varphi}A_{\Phi}$,
and comparing with~$\Braket { X'_{\rm T} |X_{\rm T} }=\delta(X'_{\rm T} - X_{\rm T})$
we obtain with~$\hbar=1$
\begin{equation}\label{A_Phi} 
|A|^2= \left ( \frac{M}{2 \pi T}\right).
\end{equation}
From this result we can write the transition probability for the pointer starting at~$X=X_0$ at~$t=0$ and subsequently evolving to the position~$X_{\rm T}$ at time~$T$. That is,
\begin{equation}\label{pointer trans prob}
|\Braket{X_0|X_{\rm T}}|^2 = \frac{M}{2\pi T},
\end{equation}
 in particular~$X_{\rm T}=X_0 + s(x,x')$.
This transition probability is independent of the initial state of the particle provided that the pointer is in an initial eigenstate of the position. Also, this relative probability is independent of the external force.  


\section{Summary and conclusion}
We have investigated the effect of a driving force acting on a harmonic oscillator during a finite von Neumann  measurement process. We have considered the general case of a pointer of finite mass and an arbitrary driving force acting on the oscillator. The coupling function in the von Neumann interaction~(\ref{Hi}) is also arbitrary. The case that has been considered in detail is that both the driving force and the coupling function are symmetric about the midpoint of the duration of the measurement.  The propagator for the system is given by the~(\ref{PropTOT}) and consists of two factors:   a factor for the pointer and a factor for the oscillator.  The factor for the pointer has the form of a free particle propagator with mass~$M_{\rm{eff}}$ given by~(\ref{Meff}), which combines the mass of the pointer and the mass of the oscillator.  In addition, the initial position of the pointer is shifted by an amount proportional to the arithmetic average of the initial and final positions of the oscillator plus a constant displacement that does not depend on the position of the particle, that is,~$s(x,x') = g(\omega,T)\left(x+x'\right)/2+d(\omega,T)$.   The constant displacement~$d(\omega,T)$ is determined by the driving force acting on the oscillator.  The factor for the oscillator consists of the propagator~(\ref{K_0}) for the free oscillator and a phase factor which is also determined by the driving force and depends on the arithmetic average of the initial and final positions of the oscillator. Thus the presence of the driving force serves to introduce a phase factor and an extra constant displacement into the propagator for the system when compared with the propagator of the system with no external force~\cite{FR}.  The entanglement between the oscillator and the apparatus is through the shift function~$s(x,x')$ that appears in the factor for the pointer in the propagator~(\ref{PropTOT}).  

When the initial  state of the oscillator is an eigenstate of the position, and the pointer is represented by a wavepacket centered at the origin, then, at the end of the measurement, the oscillator has evolved undisturbed by the interaction with the pointer and with the driving force. That is, the oscillator evolves as a free  oscillator, and the driving force has merely introduced a phase factor as shown in~(\ref{Psi_Dx_0}).  In the meantime the pointer has spread like a free particle with an effective mass~$M_\mathit{\rm{eff}}$. The center of the pointer has shifted to indicate the average between the initial and final positions of the oscillator, while the driving force has introduced an additional position independent displacement~$d(\omega,T)$.  When the initial state of the oscillator is a wavepacket,  then the final state is a superposition of sharp position states at the start of the measurement. In this case, the pointer can indicate one or another of the different positions given by the shift function~$s(x,x')$.

Unitary evolution of the system yields the~(\ref{AbsPsi_final}) for the probability distribution for the position of the oscillator at the end of the measurement. This probability distribution exhibits the entanglement between the oscillator and the apparatus through the overlap integral~(\ref{Delta_overlap}).  If the pointer is described  by a narrow wavepacket centered at the origin at the start of the measurement, then the probability distribution for the position of the oscillator at the end of the measurement is approximately the same as if the oscillator had evolved freely during the measurement. That is, neither the interaction with the pointer nor the driving force influences this result.  For the case of the oscillator starting at an eigenstate of the position, this result is exact, regardless of the initial state of the pointer, and it agrees with the result for the transition probability for the oscillator to evolve from the starting eigenstate of the position to the eigenstate of the position at time~$T$. On the other hand, if the initial state of the pointer is not narrow, then both the phase factor and the entanglement oscillator-apparatus will influence the probability distribution of the oscillator at the end of the measurement.

In addition, this probability distribution for the pointer is related to the transition probability~(\ref{trans-prob2})  for the initial position eigenstate of the oscillator to evolve to an eigenstate of the arithmetic average between the initial and final position at the end of the measurement, and also to the transition probability~(\ref{path-average2}) to an eigenstate of the average position of the oscillator~(\ref{path-average}) taken over a Feynman path.  In addition, as shown in~(\ref{pointer trans prob}), when the pointer starts in a  position eigenstate at~$X=X_0$ the transition probability to evolve to a sharp position state at the end of the measurement is uniform and depends on the ratio of the mass of the pointer to the duration of the measurement. 

To conclude, when the measurement has finite duration, both the motion of the pointer and the oscillator influence the result of the measurement. The position of the pointer at the end of the measurement  correlates with a linear combination of the initial and final positions of the oscillator.  When the coupling function in the oscillator-apparatus interaction~(\ref{Hi}) and the driving force on the oscillator are symmetric about the midpoint of the duration of the measurement, and the initial state of the  oscillator is an eigenstate of its position, the pointer will indicate the arithmetic average of the positions of the oscillator at the start and at the end of the measurement, and an additional constant displacement will appear. This additional displacement  depends on the duration of the measurement and the frequency of the oscillator and is determined by the external force acting on the oscillator.  The external force also causes the appearance of a phase factor in the wave function at the end of the measurement. The phase depends on the arithmetic average of the initial and final positions of the oscillator.  If the initial state of the pointer is a narrow wavepacket, then for any initial state of the oscillator, the measurement yields, approximately, the undisturbed probability distribution for the position of the free oscillator at the end of the measurement.  That is, the driving force plays no role in this almost perfect measurement of the position of the oscillator.

\end{document}